\documentclass[%
reprint,
amsmath,
amssymb,
aps,
prl,
nolongbibliography
]{revtex4-2}

\usepackage{graphicx}
\usepackage{dcolumn}
\usepackage{bm}

\usepackage{mathtools}

\usepackage{color}

\newcommand{\nc}{\newcommand}
\nc{\bra}{\langle}
\nc{\ket}{\rangle}
\nc{\vac}{|0\ket}
\nc{\da}{^{\dagger}}
\nc{\HASEP}{\mathcal{H}}

\nc{\so}{\hat{S}}
\nc{\nm}{\hat{n}}
\nc{\pt}{\Tilde{P}}

\nc{\im}{\text{i}}
\nc{\Imath}{\mathcal{I}}
\nc{\blue}{\textcolor{blue}}

\nc{\la}{\lambda}
\nc{\al}{\alpha}
\nc{\ze}{\zeta}

\begin{document}

\title{Exact steady states in the asymmetric simple exclusion process beyond one dimension}

\author{Yuki Ishiguro$^1$}
\author{Jun Sato$^2$}
 \affiliation{$^1$The Insitute for Solid State Physics, The University of Tokyo, 5-1-5 Kashiwanoha, Kashiwa, Chiba, 277-8581, Japan\\
 $^2$Faculty of Engineering, Tokyo Polytechnic University, 5-45-1 Iiyama-minami, Atsugi, Kanagawa 243-0297, Japan
 }

\begin{abstract}
The asymmetric simple exclusion process (ASEP) is a paradigmatic nonequilibrium many-body system that describes the asymmetric random walk of particles with exclusion interactions in a lattice. Although the ASEP is recognized as an exactly solvable model, most of the exact results obtained so far are limited to one-dimensional systems. Here, we construct the exact steady states of the ASEP with closed and periodic boundary conditions in arbitrary dimensions.
This is achieved through the concept of transition decomposition, which enables the treatment of the multi-dimensional ASEP as a composite of the one-dimensional ASEPs.
\end{abstract}
\maketitle


\textit{Introduction.}---
Exactly solvable models play a fundamental role in understanding the physics of interacting many-body systems \cite{baxterbook,faddeev1987hamiltonian,korepin1997quantum}. 
The asymmetric simple exclusion process (ASEP) is a minimal exactly solvable model for investigating interacting many-body systems far from equilibrium \cite{Derrida_1993,derrida1998exactly,blythe2007nonequilibrium,Crampe_2014,essler1996representations,golinelli2006asymmetric,gwa1992bethe,kim1995bethe,Golinelli_2004,Golinelli_2005,PhysRevE.85.042105,Motegi_2012,prolhac2013spectrum,prolhac2014spectrum,prolhac2016extrapolation,prolhac2017perturbative,ishiguro2023,de2005bethe,deGier_2006,deGier_2008,deGier_2011,Wen_2015,Crampe_2015,relation,Sandow_1994,schutz1997duality,schadschneider2000statistical,schadschneider2010stochastic,macdonald1968kinetics,klumpp2003traffic,bertini1997stochastic,PhysRevLett.104.230602,takeuchi2018appetizer}. 
The ASEP describes an asymmetric random walk of particles with exclusion interaction in a lattice.
Despite its simplicity, the ASEP captures a range of nonequilibrium phenomena, such as vehicular traffic flow \cite{schadschneider2000statistical,schadschneider2010stochastic} and biological transport phenomena \cite{macdonald1968kinetics,klumpp2003traffic}, and involves many crucial concepts in nonequilibrium physics, including the KPZ universality class \cite{bertini1997stochastic,PhysRevLett.104.230602,takeuchi2018appetizer} and boundary-induced phase transition \cite{blythe2007nonequilibrium}.
One of the most outstanding features of the ASEP is its solvability.
By employing mathematical physics approaches such as the matrix product ansatz \cite{Derrida_1993,derrida1998exactly,blythe2007nonequilibrium,Crampe_2014,essler1996representations} and the Bethe ansatz \cite{golinelli2006asymmetric,gwa1992bethe,kim1995bethe,Golinelli_2004,Golinelli_2005,PhysRevE.85.042105,Motegi_2012,prolhac2013spectrum,prolhac2014spectrum,prolhac2016extrapolation,prolhac2017perturbative,ishiguro2023,de2005bethe,deGier_2006,deGier_2008,deGier_2011,Wen_2015,Crampe_2015}, we can evaluate physical quantities without approximation.
However, most of the exact results obtained so far are limited to one-dimensional systems.

Many natural phenomena in the real world occur in systems beyond one dimension. 
For example, in highway traffic flow, roads are often not single-lane but multi-lane, and there may also be designated passing lanes. In this case, we need to consider the effect of car lane changes, which is absent in a one-dimensional system.
To understand more diverse and realistic phenomena, such as traffic flow on multi-lane highways and the dynamics of pedestrian crowds, it is vital to investigate the ASEP in multi-dimensional spaces. The extension of the ASEP to various types of two-dimensional systems has been actively studied \cite{harris2005ideal,Mitsudo_2005,pronina2006asymmetric,Reichenbach_2007,PhysRevE.76.021117,PhysRevE.77.041128,Cai_2008,jiang2009phase,xiao2009asymmetric,Schiffmann_2010,Juhász_2010,Evans_2011,Lin_2011,PhysRevE.83.031923,Yadav_2012,shi2012strong,PhysRevE.89.022131,wang2014bulk,Curatolo_2016,Klein_2016,PhysRevE.98.062111,PhysRevE.100.032133,verma2015phase,dhiman2016origin,PhysRevE.105.014128,alexander1992shock,yau2004law,singh2009transverse,PhysRevE.83.047101,PhysRevLett.119.090602,ding2018analytical,PhysRevLett.125.140601,Pronina_2004,Tsekouras_2008,PhysRevE.84.061141,Ezaki_2012,Lee_1997,PhysRevE.60.6465,wang2017dynamics,wang2018analytical}.
However, unlike the one-dimensional case, studies on the two-dimensional (2D) ASEP are mostly based on the mean-field approximation. Despite the numerous studies, exact results are few and limited to specific situations \cite{Pronina_2004,Tsekouras_2008,PhysRevE.84.061141,Ezaki_2012,Lee_1997,PhysRevE.60.6465,wang2017dynamics,wang2018analytical}.

\begin{figure}[b]
    \centering
    \includegraphics[height=7.0cm]{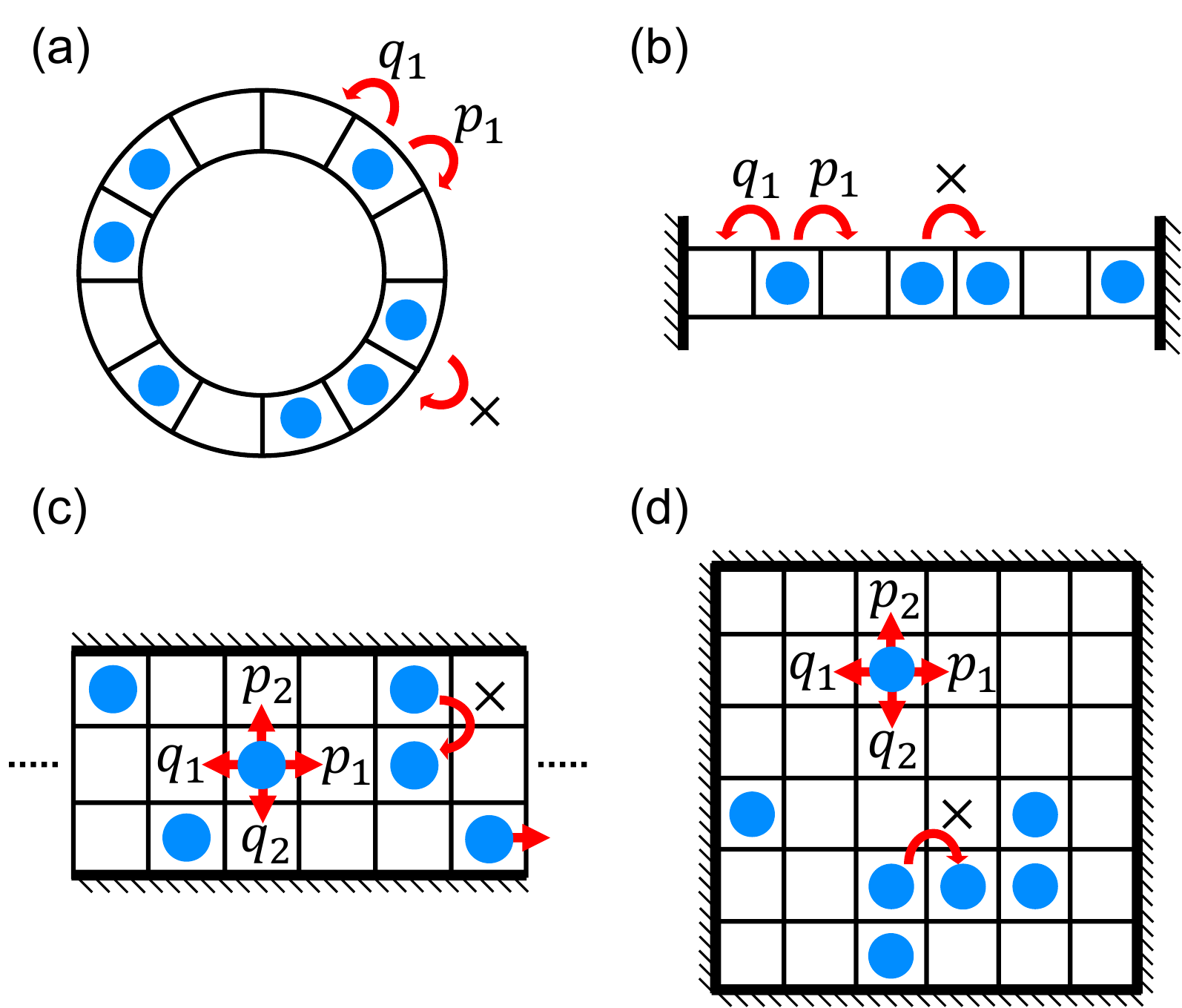}
    \caption{ASEP under various boundary conditions. The 1D ASEP with (a) periodic boundary conditions and (b) closed boundary conditions. The 2D ASEP with (c) periodic$\times$closed boundary conditions  (multi-lane ASEP) and (d) closed$\times$closed boundary conditions.}
    \label{fig:asep_fig}
\end{figure}

In this Letter, we construct the exact steady state for the ASEP in an arbitrary dimensional lattice with closed and periodic boundary conditions.
The property of the steady states depends on boundary conditions. In the one-dimensional (1D) ASEP, the density distribution in a steady state is spatially homogeneous in periodic boundary conditions (Fig. \ref{fig:asep_fig}(a)) \cite{blythe2007nonequilibrium}, while that is inhomogeneous in the closed boundary conditions  (Fig. \ref{fig:asep_fig}(b)) \cite{Sandow_1994}.
In higher dimensions, there are more patterns of boundary conditions. Here, we consider the combinations of periodic and closed boundary conditions.
In the two-dimensional case, there are three types of combinations depending on the choice of boundary conditions for the horizontal and vertical directions (Fig. \ref{fig:asep_fig}(c)-(d))): (1) periodic$\times$periodic boundary conditions (torus), (2) periodic$\times$closed boundary conditions (multi-lane ASEP), and (3) closed$\times$closed boundary conditions. 
In the following, we demonstrate that steady states corresponding to such various boundary conditions can be exactly constructed in any dimension.

\textit{Model.}---
The ASEP is a continuous-time Markov process that describes the asymmetric diffusion of particles with hardcore interactions. 
The ASEP is usually considered in a one-dimensional lattice.
The updating rule of the 1D ASEP is defined as follows.
Each particle moves to the nearest forward (backward) site with a hopping rate $p$ $(q)$. Due to the exclusion interactions, each site can contain at most one particle.
In periodic boundary conditions (Fig. \ref{fig:asep_fig}(a)), a particle at site $1$ ($L$) hops to site $L$ ($1$) with rate $q$ ($p$) and to site $2$ ($L-1$) with rate $p$ ($q$).
On the other hand, in closed boundary conditions (Fig. \ref{fig:asep_fig}(b)), a particle at site $1$ ($L$) hops only to site $2$ ($L-1$) with rate $p$ ($q$). 

In this study, we consider the multi-dimensional ASEP, whose updating rule is given below.
We consider a $d$-dimensional lattice with the system size $L_1\times L_2 \times \cdots \times L_d$.
Each particle moves to the nearest forward (backward) site in the $i$ direction ($1\le i \le d$) with a hopping rate $p_i$ ($q_i$).
Each site can contain at most one particle because of the exclusion interactions. For a given $\ell$ ($0\le \ell \le d$), we consider closed boundary conditions for $i\le \ell$ direction and periodic boundary conditions for $i\ge \ell+1$ direction.

The position of a site $\bm{r}$ is denoted as $\bm{r}=(r_1,r_2,\cdots,r_d)$ ($1\le r_i \le L_i$ for $1\le i\le d$).
The state of a site $\bm{r}$ is represented by a Boolean number $n_{\bm{r}}$, which is set to $n_{\bm{r}}=0$ ($n_{\bm{r}}=1$) if the site is empty (occupied). 
A configuration $n$ is described by a series of the Boolean numbers $(n_{(1,\cdots,1)},n_{(1,\cdots,2)},\cdots,n_{(L_1,\cdots,L_d)})$. 
We denote the probability of the system being in a configuration $n$ at time $t$ as $P(n,t)$.
The time evolution of $P(n,t)$ is determined by the master equation
\begin{align}
\begin{split}    
    &\frac{d}{dt} P(n,t) \\
    &=\sum_{n'\neq n} \left[ P(n',t) W(n' \to n)-P(n,t) W(n \to n') \right] \\
    &=\sum_{n' \in I_n} P(n',t) W(n' \to n) - \sum_{n'\in D_n} P(n,t) W(n \to n'),
\end{split}
\label{eq:mastereq}
\end{align}
where $W(n \to n')$ is a transition rate from a configuration $n$ to $n'$, $I_n$ represents a set of all configurations that can transition to a configuration $n$, and $D_n$ denotes a set of all configurations that can transition from a configuration $n$.

It is helpful to express the master equation (\ref{eq:mastereq}) in vector form. The state of a site $\bm{r}$ is described by a two-dimensional vector $|n_{\bm{r}}\ket$, which equals $|0\ket$ for empty and $|1\ket$ for occupied. 
A configuration $n$ is represented by $|n\ket= \bigotimes_{r_1=1}^{L_1}\bigotimes_{r_2=1}^{L_2}\cdots\bigotimes_{r_d=1}^{L_d}|n_{\bm{r}}\ket$, which forms an orthonormal basis of the configuration space under normalization.
A stochastic state vector $|P(t)\ket$ is described by 
\begin{align}
    |P(t)\ket = \sum_{n} P(n,t)|n\ket.
    \label{eq:statevector}
\end{align}
The master equation (\ref{eq:mastereq}) is given by
\begin{align}
    \frac{d}{dt}|P(t)\ket =\HASEP|P(t)\ket,
    \label{eq:imschrodinger}
\end{align}
where $\HASEP$ is the Markov matrix, which is expressed as a non-Hermitian spin chain as follows
\begin{align} 
\begin{split}
    \HASEP 
    =&\left( \sum_{i=1}^{\ell}\sum_{r_i=1}^{L_i-1}\sum_{\{r_1,\cdots,r_d\} \setminus r_i} 
    +\sum_{i=\ell+1}^{d}\sum_{\{r_1,\cdots,r_d\}} 
    \right)\\
    &\qquad \left[ p_i \left\{ \so_{\bm{r}}^{+}\so_{\bm{r}+\bm{e}_i}^{-}- \nm_{\bm{r}}\left(1-\nm_{\bm{r}+\bm{e}_i}\right) \right\} \right. \\
    & \qquad \left. + q_i \left\{ \so_{\bm{r}}^{-}\so_{\bm{r}+\bm{e}_i}^{+}- \left(1-\nm_{\bm{r}}\right) \nm_{\bm{r}+\bm{e}_i}\right\} \right].
\end{split}
\label{eq:ASEP}
\end{align}
Here, we consider the half of the Pauli matrices $\so_{\bm{r}}^{x,y,z}$ that act on a site $\bm{r}$, and introduce the ladder operators $\so_{\bm{r}}^{\pm}=\so_{\bm{r}}^{x}\pm \im \so_{\bm{r}}^{y}$ and the number operators $\nm_{\bm{r}}=I/2-\so_{\bm{r}}^z$. $\bm{e}_i$ represents the unit vector in $i$ direction.
The sum range corresponds to the boundary conditions. The first sum represents closed boundary conditions, and $\sum_{\{r_1,\cdots,r_d\} \setminus r_i}=\sum_{r_1=1}^{L_1}\cdots \sum_{r_{i-1}=1}^{L_{i-1}}\sum_{r_{i+1}=1}^{L_{i+1}}\cdots \sum_{r_d}^{L_d}$.
The second represents periodic boundary conditions, and $\sum_{\{r_1,\cdots,r_d\}}=\sum_{r_1=1}^{L_1}\cdots \sum_{r_d=1}^{L_d}$.
When the hopping rates are symmetric, the Hamiltonian of the ASEP is equivalent to that of the spin-$1/2$ Heisenberg model. In this sense, the ASEP is regarded as a non-Hermitian extension of the Heisenberg model by introducing asymmetricity.

\textit{Steady state.}---
In the AESP, any initial state relaxes to the steady state.
Since the number of particles $N$ is conserved in closed and periodic boundary conditions, the Hamiltonian is block diagonalized, and one steady state exists for each $N$.
In the following, we fix the particle number $N$ and consider a $\binom{L_1 \cdots L_d}{N}$-dimensional subspace.
We denote the position of an $i$-th particle ($1\le i \le N$) as $\bm{r}_i=(r_{i;1},r_{i;2},\cdots,r_{i;d})$. 
A configuration $n$ is represented by a set of the $N$ partile positions $\{\bm{r}_1,\bm{r}_2,\cdots,\bm{r}_N\}$.

The probability distribution in a steady state $P_{st}(n)$ is the solution
of the master equation (\ref{eq:mastereq}) with $\frac{d}{dt} P(n,t) =0$:
\begin{align}
   \sum_{n' \in I_n} P_{st}(n') W(n' \to n) - \sum_{n'\in D_n} P_{st}(n) W(n \to n')=0
    \label{eq:st_mastereq}
\end{align}
In vector form, the steady state $|P_{st}\ket$ corresponds to the eigenvector of the Markov matrix (\ref{eq:ASEP}) with an eigenvalue zero
\begin{align}
    \HASEP |P_{st}\ket =0.
\end{align}
We express the steady state vector $|P_{st}\ket$ as
\begin{align}
    |P_{st}\ket = \frac{1}{Z}\sum_n \pt_{st}(n) |n\ket,
    \label{eq:stvector}
\end{align}
where $\pt_{st}(n)$ is the weight of the probability distribution that satisfies $P_{st}(n)=\pt_{st}(n)/Z$, and $Z=\sum_n \pt_{st}(n)$ is the normalization constant to satisfy $\sum_{n}P_{st}(n)=1$. Note that $\pt_{st}(n)$ also satisfies the master equation for a steady state (\ref{eq:st_mastereq}).

In the one-dimensional case, the exact steady state is constructed for both periodic \cite{blythe2007nonequilibrium} and closed boundary conditions \cite{Sandow_1994}.
The weight of the probability distribution for a steady state under periodic boundary conditions is given by 
\begin{align}
    \pt_{st,p}(n)=1 \quad \text{for } \forall n.
    \label{eq:weight_pbc}
\end{align}
From the master equations (\ref{eq:st_mastereq}), the following relation is satisfied
\begin{align}
    \sum_{n' \in I_{n}} W_{p}(n' \to n) -\sum_{n'\in D_{n}} W_{p}(n \to n')=0,
    \label{eq:rel_periodic_st}
\end{align}
where $W_p(n \to n')$ represents the transition probability in the 1D periodic ASEP.
On the other hand, the weight of the probability distribution for a steady state under closed boundary conditions is given by 
\begin{align}
    \pt_{st,c}(n)=\left(\frac{p_1}{q_1}\right)^{\sum_{j=1}^N r_{j;1}}.
\end{align}
Based on the master equation (\ref{eq:st_mastereq}), we obtain
\begin{align}
\begin{split}    
    &\sum_{n' \in I_{n}}\left(\frac{p_1}{q_1}\right)^{\sum_{j=1}^N r'_{j;1}} W_{c}(n' \to n) \\
    &-\sum_{n'\in D_{n}} \left(\frac{p_1}{q_1}\right)^{\sum_{j=1}^N r_{j;1}} W_{c}(n \to n')=0,
\end{split}
\label{eq:rel_closed_st}
\end{align}
where $W_c(n \to n')$ denotes the transition probability in the 1D closed ASEP.

\begin{figure}[tbh]
    \centering
    \includegraphics[height=8.0cm]{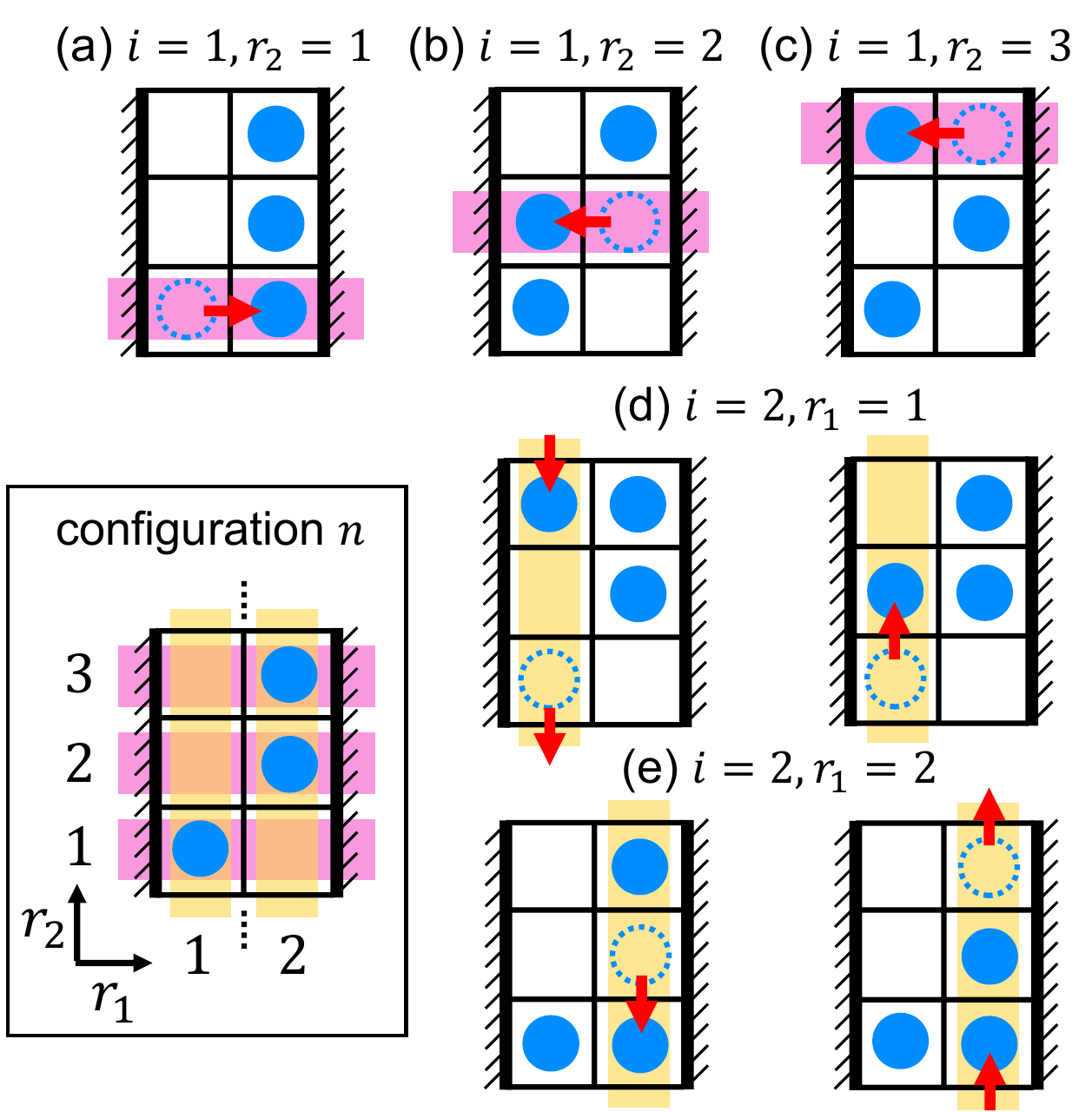}
    \caption{Example of the transition of states from a configuration $n=\{ (1,1), (2,2), (2,3) \}$ in the 2D ($2\times3$) ASEP with $\ell=1$. All configurations that can transition from a configuration $n$ in the 2D ASEP can be regarded as the transition of the five 1D ASEPs (a)-(e).}
    \label{fig:decomp_fig}
\end{figure}

In this Letter, we construct the exact steady state for the multi-dimensional ASEP with closed and periodic boundary conditions (\ref{eq:ASEP}).
The weight of the probability distribution for a steady state is given by
\begin{align}
    \pt_{st,m}(n)=\prod_{i=1}^{\ell}\left(\frac{p_i}{q_i}\right)^{\sum_{j=1}^N r_{j;i}}.
    \label{eq:weight_steady}
\end{align}
In the following, we show that this gives a stationary solution to the master equation (\ref{eq:st_mastereq}). 
The key is the decomposition of the multi-dimensional ASEP to the 1D ASEP. 
All transitions of a configuration in the multi-dimensional ASEP can be regarded as those of the corresponding 1D ASEP.
Fig. \ref{fig:decomp_fig} shows, as an example, the transition from a configuration $n=\{ (1,1), (2,2), (2,3) \}$ in the 2D ($2\times3$) ASEP with $\ell=1$. 
In this case, we find two 1D ASEPs extending in the $1$-direction ($r_2=1, 2$) and three 1D ASEPs extending in the $2$-direction ($r_1=1,2,3$). By considering the configuration transition in each 1D ASEP (Fig, \ref{fig:decomp_fig}(a)-(e)) and aggregating all these transitions, we obtain all configurations that can transition from the configuration $n$ in the 2D ASEP.

When the coordinates $\{r_1,\cdots,r_d\}$ except $r_i$ (denote them as $\{r\}\setminus r_i:=\{r_1,\cdots,r_d\} \setminus r_i$) are fixed, we can specify a series of cells arranged on a one-dimensional line extending $i$ direction. 
Regarding the cells as the 1D ASEP, we consider the transition of states for a given configuration $n$. 
We denote a set of all configurations that can transition to (from) a configuration $n$ in the 1D ASEP extending $i$ direction with fixed $\{r\} \setminus r_i$ as $I^{i}_{n;\{r\} \setminus r_i}$ ($D^{i}_{n;\{r\} \setminus r_i}$).
The decomposition of the multi-dimensional ASEP to the 1D ASEP means that the configuration set $I_n$ ($D_n$) for the multi-dimensional ASEP can be expressed as the sum of the sets $I^{i}_{n;\{r\} \setminus r_i}$ ($D^{i}_{n;\{r\} \setminus r_i}$) for the 1D ASEPs
\begin{align}
    I_n = \bigcup_{i=1}^d\bigcup_{\{r\} \setminus r_i} I^{i}_{n;\{r\} \setminus r_i}, \quad D_n = \bigcup_{i=1}^d\bigcup_{\{r\} \setminus r_i} D^{i}_{n;\{r\} \setminus r_i}.
\end{align}

Based on this picture, we decompose the master equations of the multi-dimensional ASEP (\ref{eq:mastereq})
\begin{align}
\begin{split}
    &\frac{d}{dt}P(n,t) \\
    &=\sum_{i=1}^{d} \sum_{\{r\}\setminus r_i} \left[\sum_{n' \in I^{i}_{n;\{r\} \setminus r_i}} P(n',t) W^{i}_{n;\{r\} \setminus r_i}(n' \to n) \right.\\
    &\left. \hspace{1.5cm} -\sum_{n'\in D^{i}_{n;\{r\} \setminus r_i}} P(n,t) W^{i}_{n;\{r\} \setminus r_i}(n \to n') \right],
\end{split}
\label{eq:d_mastereq}
\end{align}
where $W^{i}_{n;\{r\} \setminus r_i}(n \to n')$ represents a transition rate from a configuration $n$ to $n'$ in the 1D ASEP extending $i$ direction with fixed $\{r\}\setminus r_i$. Then, we substitute Eq. (\ref{eq:weight_steady}) for the right-hand side of the master equation (\ref{eq:d_mastereq})
\begin{align}
\begin{split}
    &\sum_{i=1}^{d} \sum_{\{r\}\setminus r_i} \left[\sum_{n' \in I^{i}_{n;\{r\} \setminus r_i}} \pt_{st,m}(n',t) W^{i}_{n;\{r\} \setminus r_i}(n' \to n) \right.\\
    &\left. \hspace{1.5cm} -\sum_{n'\in D^{i}_{n;\{r\} \setminus r_i}} \pt_{st,m}(n,t) W^{i}_{n;\{r\} \setminus r_i}(n \to n') \right] \\
    &=\sum_{i=1}^{\ell} \sum_{\{r\}\setminus r_i} \prod_{k \neq i}^{\ell} \left(\frac{p_k}{q_k}\right)^{\sum_{j=1}^N r_{j;k}} \\
    &\hspace{0.5cm} \left[\sum_{n' \in I^{i}_{n;\{r\} \setminus r_i}} \left(\frac{p_i}{q_i}\right)^{\sum_{j=1}^N r'_{j;i}} W^{i}_{n;\{r\} \setminus r_i}(n' \to n) \right.\\
    &\left. \hspace{0.5cm} -\sum_{n'\in D^{i}_{n;\{r\} \setminus r_i}} \left(\frac{p_i}{q_i}\right)^{\sum_{j=1}^N r_{j;i}} W^{i}_{n;\{r\} \setminus r_i}(n \to n') \right] \\
    &+\sum_{i=\ell+1}^{d} \sum_{\{r\}\setminus r_i} \prod_{i =1}^{\ell} \left(\frac{p_i}{q_i}\right)^{\sum_{j=1}^N r_{j;i}} \\
    &\hspace{0.5cm} \left[\sum_{n' \in I^{i}_{n;\{r\} \setminus r_i}} W^{i}_{n;\{r\} \setminus r_i}(n' \to n) \right.\\
    &\left. \hspace{2.5cm} -\sum_{n'\in D^{i}_{n;\{r\} \setminus r_i}}  W^{i}_{n;\{r\} \setminus r_i}(n \to n') \right] \\
    &=0.
\end{split}
\end{align}
Here, we use Eq. (\ref{eq:rel_periodic_st}) and Eq. (\ref{eq:rel_closed_st}), which are the relation of the steady state for the 1D ASEP.
Therefore, Eq. (\ref{eq:weight_steady}) is a stationary solution of the master equation.

Here, it is worth mentioning that in the case of periodic boundary conditions, it can be extended to non-uniform lanes. 
In other words, even if the hopping rates in $i$ directions ($\ell+1 \le i \le d$) are extended to depend on coordinates $\{r_1, \cdots, r_d\}$ except $r_i$ (that is, $p_i=p_i(\{r\}\setminus r_i)$, $q_i=q_i(\{r\}\setminus r_i)$), Eq. (\ref{eq:weight_steady}) still gives a stationary solution of the master equation (\ref{eq:st_mastereq}). 
Since the weight of the stationary probability distribution under periodic boundary conditions (\ref{eq:weight_pbc}) is independent of the configuration, we can show this through a parallel discussion.

\begin{figure}[tb]
    \centering
    \includegraphics[height=8.0cm]{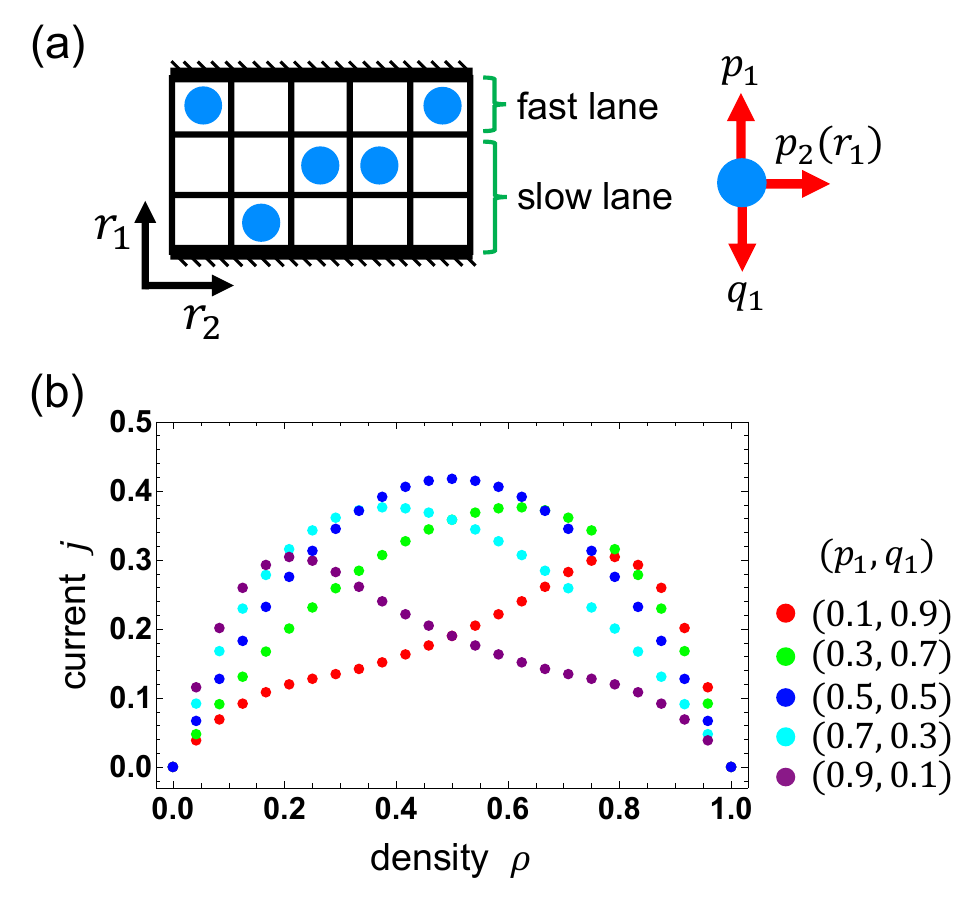}
    \caption{(a) 2D ASEP with $\ell=1$. We set $(L_1,L_2)=(3,8)$, $p_2(r_1)=0.3$ for $r_1=1,2$ and $p_2(r_1)=1.0$ for $r_1=3$, and $q_2(r_2)=0$. (b) Relation between the quasi-one-dimensional current $j$ and the density $\rho$ for various hopping rates $(p_1,q_1)$ in the 2D ASEP.}
    \label{fig:example}
\end{figure}

\textit{Example.}---
In the following, as an example, we consider the quasi-one-dimensional flow in the 2D ASEP with $\ell=1$ (Fig. \ref{fig:example}(a)) and show the effect of two-dimensionality.
We introduce two types of lanes (fast lane and slow lane) by setting inhomogeneous hopping rates in the $2$ direction ($p_2 (r_1)$). For simplicity, we assume $q_2 (r_1)=0$.
From Eq. (\ref{eq:weight_steady}), the steady state of the ASEP is given by 
$|P_{st}\ket = \frac{1}{Z}\sum_{n}\left(\frac{p_1}{q_1}\right)^{\sum_{j=1}^N r_{j;1}} |n\ket$. The expectation value of a physical quantity $\hat{A}$ in the steady state is expressed as $\bra \hat{A} \ket = \bra s | \hat{A} | P_{st}\ket$ where $\bra s |= \sum_{n} \bra n |$. Here, we introduce a quasi-one-dimensional current operator in the $2$ direction at the cross-section $r_2=k$ as 
\begin{align}
\begin{split}    
    \hat{j}_{k}=\sum_{r_1=1}^{L_1}  p_2(r_1) \bra \hat{n}_{(r_1,k)} (1-\hat{n}_{(r_1,k+1)}) \ket 
\end{split}
\end{align}
and define a quasi-one-dimensional current as $j=\bra \hat{j}_1 \ket$.
Then, the current is given by
\begin{align}
    j = \frac{1}{Z}\sum_{r_1=1}^{L_1}\sum_{\Tilde{n}} p_2(r_1)\left( \frac{p_1}{q_1} \right)^{\sum_{j=1}^N r_{j;1}},
\end{align}
where $\sum_{\Tilde{n}}$ represents the sum over all configurations with $n_{(r_1,1)}=1$ and $n_{(r_1,2)}=0$.

Fig. \ref{fig:example}(b) shows the relation between the quasi-one-dimensional current $j$ and the density $\rho=\frac{N}{L_1L_2}$ for various hopping rates in the $1$ direcrtion $(p_1,q_1)$.
In the case of symmetric rates $p_1=q_1$ (blue dots), two-dimensionality does not significantly affect, and the relation is almost equivalent to that in the 1D ASEP. Namely, the current $j$ reaches its maximum value when the density $\rho$ equals $1/2$.
Conversely, when the hopping rate is asymmetric $p_1 \neq q_1$,
two-dimensionality alters the properties of the flow. Specifically, the value of the density $\rho$ at which the current $j$ reaches its maximum deviates from $1/2$.

\textit{Conclusion.}---
In this Letter, we presented the exact results of the ASEP in more than one dimension. 
We introduced the multi-dimensional ASEP with closed and periodic boundary conditions, which describes a range of situations, such as asymmetric diffusion in a box and quasi-one-dimensional flow in a tube. 
We constructed the exact steady states of the ASEP in arbitrary dimensions, and, as an example, we revealed the effect of two-dimensionality on quasi-one-dimensional flow by calculating the current in the 2D inhomogeneous ASEP with slow and fast lanes.
The central concept was the decomposition of transitions, which enabled us to treat the multi-dimensional ASEP as the combination of the 1D ASEPs.
This would be applicable to investigate other exactly solvable models in higher dimensions.






\bibliography{apssamp}

\end{document}